\documentclass[jcp, floatfix, nobibnotes, reprint, superscriptaddress]{revtex4-1}
\usepackage{docs}%
\usepackage{siunitx}
\usepackage{amsmath}
\DeclareSIUnit[number-unit-product = {\,}]
\cal{cal}
\DeclareSIUnit\kcal{\kilo\cal}
\DeclareSIUnit[number-unit-product = {\,}]
\Btu{Btu}
\DeclareSIUnit[number-unit-product = {\,}]
\Fahr{\degree F}
\DeclareSIUnit[number-unit-product = {\,}]
\usepackage{bm}%
\usepackage[colorlinks=true,linkcolor=blue]{hyperref}%
\usepackage{graphicx}
\usepackage[utf8]{inputenc}
\usepackage{tabularx}
\usepackage{dcolumn}
\usepackage{epstopdf}
\usepackage{afterpage}
\usepackage{setspace}
\usepackage{multirow}
\usepackage{dsfont}
\usepackage{sidecap}
\usepackage{listings}
\usepackage{xcolor} 
\definecolor{codegreen}{rgb}{0,0.6,0}
\definecolor{codegray}{rgb}{0.5,0.5,0.5}
\definecolor{codepurple}{rgb}{0.58,0,0.82}
\definecolor{backcolour}{rgb}{0.95,0.95,0.92}
\lstdefinestyle{mystyle}{
    backgroundcolor=\color{backcolour},   
    commentstyle=\color{codegreen},
    keywordstyle=\color{magenta},
    numberstyle=\tiny\color{codegray},
    stringstyle=\color{codepurple},
    basicstyle=\ttfamily\footnotesize,
    breakatwhitespace=false,         
    breaklines=true,                 
    captionpos=b,                    
    keepspaces=true,                 
    numbers=left,                    
    numbersep=5pt,                  
    showspaces=false,                
    showstringspaces=false,
    showtabs=false,                  
    tabsize=2
}

\usepackage{booktabs} 
\DeclareMathAlphabet\mathbfcal{OMS}{cmsy}{b}{n}

\newcommand{\alsoaffiliation}[1]{\affiliation{#1}}
\setcitestyle{super}
\newcommand{\nbits}{n_{\mathrm{bits}}}
\newcommand{\slopeofslopes}{\tilde{m}^{P}}
\newcommand{\interceptofslopes}{\tilde{c}^{P}}

\begin{document}

\title{Exact sampling of molecules in chemical space}

\author{Jan Weinreich}

\affiliation{University of Vienna, Faculty of Physics, Kolingasse 14-16, AT-1090 Wien, Austria}
\affiliation{University of Vienna, Vienna Doctoral School in Physics, Boltzmanngasse 5, 1090 Vienna, Austria}
\author{Konstantin Karandashev}
\affiliation{University of Vienna, Faculty of Physics, Kolingasse 14-16, AT-1090 Wien, Austria}
\author{Guido Falk von Rudorff}
\affiliation{University Kassel, Department of Chemistry, Heinrich-Plett-Str.40, 34132 Kassel, Germany}
\alsoaffiliation{Center for Interdisciplinary Nanostructure Science and Technology (CINSaT), Heinrich-Plett-Straße 40, 34132 Kassel}
\email{vonrudorff@uni-kassel.de}

\date{\today}

\begin{abstract}
The concept of molecular similarity appears in many machine-learning algorithms based on the assumption that molecules with similar representations will also share similar properties. In this work, we propose a new way to study similarity measures in molecular graph space using a Monte Carlo approach.
We enable direct sampling from the underlying distribution of chemical space without numerical approximations or complete enumeration of molecular graphs, the latter intractable for practically relevant graph sets of interest. The Monte Carlo method allows observation of several interesting fundamental properties of chemical space, such as a linear trend of average property derivatives in chemical space with respect to the property's value at the molecule of interest. The trend was observed for extensive and intensive properties, suggesting that this trend is an inherent property of chemical space.
\end{abstract}

\maketitle 

\bigskip

\section{Introduction}

Chemical space\cite{kirkpatrick2004chemical} encompasses all molecules occurring naturally, synthesized, or existing only in theory. A key concept to study this vast combinatorial space is molecular similarity\cite{Johnson1990ConceptsAA}. It is essential for techniques that assume that more similar molecules tend to have more similar properties, such as any similarity-based machine learning (ML) method or proposing experimentally accessible molecules.\cite{Olivecrona2017Molecular} For similarity-based machine learning, there is a large body of research on defining similarity to improve predictive accuracy. Including physical knowledge of a property of interest has proven a particularly viable approach.\cite{10.1093/pnasnexus/pgac039, Goscinski_2021, C6CP00415F,Parsaeifard_2021, doi:10.1021/acs.chemrev.1c00021}
For instance, consider a representation that does not explicitly include orbital information. This omission can lead to a situation where molecules in the close neighborhood or even at zero distance are very distinct in terms of their HOMO-LUMO gap values.\cite{D2MA00742H,D1DD00050K, 10.1063/5.0151122} This results in large prediction errors due to the injectivity problem and the uniqueness requirement of representations.\cite{von2013representation} Hence, rationalizing the choice of the induced molecular neighborhood in relation to the selected representation and metric is essential for accurate predictions.

Here our focus is on numerically studying fundamental properties of neighborhoods\cite{neighbourhood2013} in chemical space, \emph{i.e.} given a metric and representation, we ask: \textit{What are the molecules surrounding any given molecule in chemical space?} This task can also be formulated as studying \emph{topologies}, or, more exactly, the set of all open \emph{balls}, generated for chemical space using a given similarity measure.\cite{munkres2018topology, KORN2023102578}
Unfortunately, a simple enumeration of all molecules in the neighborhood of every molecule becomes unfeasible for molecules over a large size, due to the immense combinatorial complexity that arises from the diverse combinations of compositions and molecular graphs.
Our goal is to study neighborhoods of molecules in CCS exactly without resorting to exhaustive enumeration. Instead, we introduce an exact sampling of distributions defined in terms of molecular similarity. We address the question of which chemical neighborhood results from a specific representation and metric.
The latter is achieved with an Evolutionary Monte Carlo (EMC) algorithm\cite{Liang_Wong:2000} proposed recently\cite{mosaics}.
The work presented in~\citenum{chemspaceold} is similar in spirit as it includes random cross-over and mutation moves but does not allow exact sampling of the distance distribution.
\begin{figure}[htb]
          \centering           
    \includegraphics[width=0.8\columnwidth]{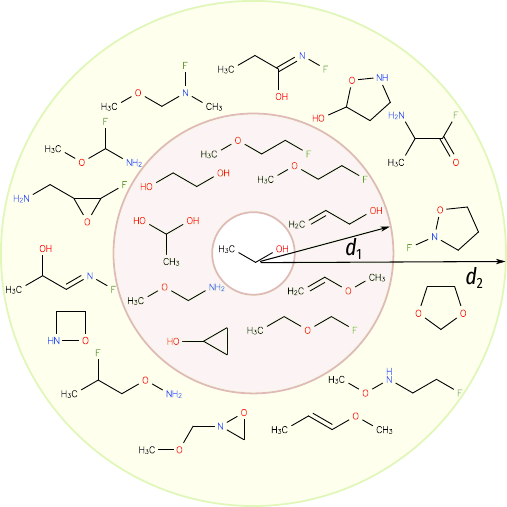}
          \caption{
          Illustration of the molecules in the environment of ethanol in the representation space. The first shell contains neighbors up to a distance of $d_1$. A second shell contains molecules in the interval $d_{1} \leq d \leq d_{2}$.}
     \label{fig:radar} 
 \end{figure}
The study by \textcite{Tibo2023Exhaustive} focused on an approximate topology resulting from a vast database of drug-like molecules. Rather than aiming for approximately \textit{exhaustive} sampling as in the latter, we focus on \textit{exact representative} sampling with analytic representation vectors.\cite{doi:10.1021/acs.chemrev.1c00021, soap} Several effective search algorithms for identifying molecules in the representation environment from a large database were proposed\cite{WILLETT20061046, spotlight}. By contrast, our approach allows for the generation of molecules based only on defined distance distributions. In other earlier studies\cite{ Awale_Visini_Probst_Arus-Pous_Reymond_2017, map4}, the focus is on visualizing and understanding the structure of the chemical space topology itself. Several methods\cite{D1SC00231G} are capable of sampling the chemical environment by modifying the molecular graphs \textit{i.e.} generating random molecules that are somewhat similar to a given molecule is a solved problem. We also note that this paper's theme of sampling exactly without complete enumeration is shared with Ref.~\cite{doi:10.1021/acs.jcim.4c00147} , which was solving the same problem for Chemical Fragment Spaces.

However, to the best of our knowledge, none precisely sample the distribution in terms of distances to systematically expand chemical space around a given molecule and representation. Our method achieves this without any bias or machine learning-based training -- the target molecule is enough.

We note that our work could be used to investigate other problems defined in terms of molecular similarities. For example, finding activity cliffs\cite{doi:10.1021/acs.jcim.2c01073} by scanning the molecules in the chemical environment and identifying the closest points with large property differences. Such applications would not be fundamentally different from minimization problems discussed in Ref.~\citenum{mosaics}, for which EMC is guaranteed to eventually find the exact solution thanks to the properties of Markov chains. However, such problems were beyond the scope of this work in which we wish to introduce the methodology. First, we introduce potential functions defined in CCS and present details of the algorithm (Sec.~\ref{sec:details}). Then we discuss results based on the sampling of the chemical environment of molecules (Sec.~\ref{sec:results}). Finally, we analyze an identified relationship between property values and the magnitude of derivatives of a property with respect to representation distances (Sec.~\ref{sec:relating_d_and_p}).

\section{Methods}
\begin{figure*}[htb]
          \centering           
            \includegraphics[width=0.8\linewidth]{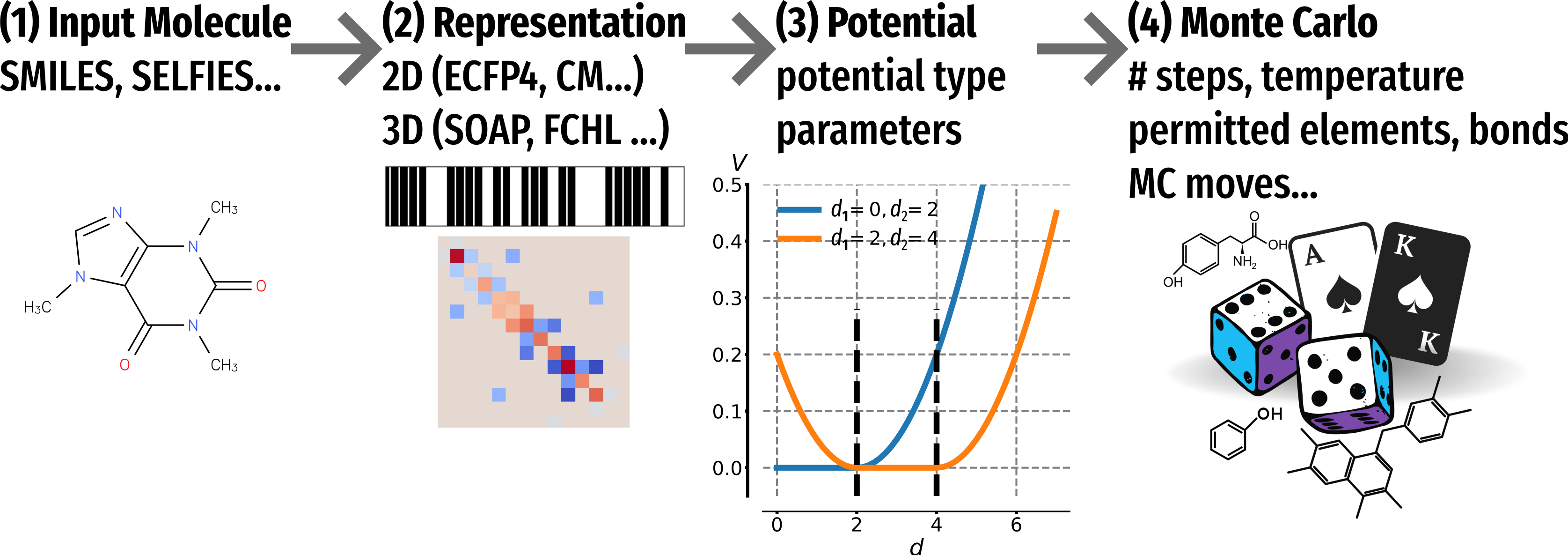}
        \caption{\texttt{ChemSpaceSampler}: (1) central molecule, (2) 2D or 3D representations, and (3) potential function, s. Eq.~(\ref{eq:flat_parabola}), are defined. The (4) MC simulation is initialized for a given number of steps, temperatures for the replica exchanges as well as permitted elements and bonds, and number of heavy atoms. For (3), the potential function, two quadratic potential functions are shown, blue potential with $d_1=0$ and $d_2=2$, orange curve with $d_1=2$ and $d_2=4$.
        }
     \label{fig:components_code} 
\end{figure*}
\subsection{Neighborhoods in chemical space}
\label{sec:details}

Our goal is to obtain insights into the structure of chemical space from an exact sampling of \emph{neighborhoods} in chemical space within \emph{shells} defined in terms of similarity to a central molecule (see Fig.~(\ref{fig:radar})). Here we focused on similarity measures expressed in terms of distances between two molecular graphs $A$ and $B$, defined as
\begin{align}
d(A, B) = || \mathbf{X}(A) - \mathbf{X}(B) ||_{2},
\end{align}
where $\mathbf{X}$ is a representation vector and $||\ldots||_{2}$ is the Euclidean norm. Smaller values of $d(A,B)$ mean greater similarity between $A$ and $B$. Other metrics\cite{caylak_2020,Fabregat_2022} or similarity expressions based on kernel functions could be used as well. 
Here, we select two choices for representations that can be computed from the molecular graph or geometry but we note that embedding vectors of neural networks \textit{e.g.} from SchNet\cite{schnet} may also be used:
\begin{enumerate}
    \item ECFP4\cite{ecfp} fingerprint, a group-based representation with bit resolution $\nbits$, unless specified otherwise, $\nbits=2048$. The diameter parameter, here four, is the maximum distance of neighborhoods of each atom.
    \item 
    Smooth overlap of atomic positions\cite{soap} (SOAP) a ``physics-based'' representation\cite{parsaeifard2021}  averaged over atomic sites,\cite{dscribe} for the remainder of the text referred to as SOAP. 3D conformations are generated with the Morfeus\cite{Kjell2022} package using MMFF94 forcefield,\cite{Halgren:1996_I,*Halgren:1996_II,*Halgren:1996_III,*Halgren_Nachbar:1996_IV,*Halgren:1996_V,*Halgren:1999_VI,*Halgren:1999_VII, Kjell2022} as implemented\cite{Tosco_Landrum:2014} in RdKit,\cite{software:RDKit} and the recipe from Ref.~\citenum{Ebejer_Deane:2012}. Conformer-averaging of the representation accounts for the variance of the distance within the classical Boltzmann ensemble for a given temperature.\cite{fml_paper}
\end{enumerate}

An average over a shell or a neighborhood can be estimated by randomly drawing sets of molecules drawn from a distribution that is uniform over all such molecules $B$ that $d_1 \leq d(A_{\mathrm{c}},B) \leq d_2$, where $A_{\mathrm{c}}$ is a \emph{central molecule}. Simulating such a distribution directly with Monte Carlo (MC), however, risks generating trajectories confined to disconnected subsets of molecular graphs in the neighborhood. To create a simulation protocol that avoids the issue, we define a potential function in terms of distance to the central molecule (from now on $d$ for short)
\begin{align}
V(d) = 
\begin{cases}
V_{0}(d-d_1)^2  &\text{if}\ d < d_1 \\
0 &\text{if}\ d_1 \leq d \leq d_2 \\
V_{0}(d-d_2)^2  &\text{if}\ d > d_2~.
\end{cases}
\label{eq:flat_parabola}
\end{align}
The potential is based on an absolute measure of similarity and predefined representation functions and must be adapted to the specific chemical space of interest.

Similar to parallel tempering,\cite{Hukushima_Nemoto:1996,Sambridge:2014,Angelini_Ricci-Tersenghi:2019} several replicas are introduced with corresponding temperatures $\beta_{i}$ generating a trajectory whose probability density converges to the Boltzmann distribution with unnormalized distribution function $\rho_{i}$
\begin{equation}
\rho_{i}(d)=\exp[-\beta_{i}V(d)].
\end{equation}
The potential $V(d)$ has a flat plateau depicted in Fig.~\ref{fig:components_code}, with the resulting $\rho_{i}(d)$ distributions biased towards $d_1 \leq d \leq d_2$. $\beta_{i}$ include infinitely large values (see \emph{greedy replicas} in Ref.~\citenum{mosaics}) that can only accept trial moves to graphs with potential functions smaller or equal to the current one. Therefore, as the simulation progresses for each greedy replica the potential decreases until the replica reaches the plateau. Once it is inside the plateau it can never leave it by definition. The presence of finite values of $\beta_{i}$ (see \emph{exploration replicas} in Ref.~\citenum{mosaics}) ensures our MC simulations do not get obstructed in moving between subsets of chemical space separated by barriers of $V$.

Sometimes it is preferable to limit the sampling to molecules that satisfy conditions related to stability or synthesizability.\cite{skoraczynski2023critical} For example when a machine learning approach should be designed to only include molecules that meet these conditions.

Such examples were beyond the scope of this research, but we note that constraints can be accommodated by adding a non-negative term to $V$~Eq.~(\ref{eq:flat_parabola}) which is $0$ if and only if the considered chemical graph satisfies said constraint.

\subsection{Numerical details}

As illustrated in Fig.~\ref{fig:components_code}, the central molecule $A_c$ is first defined (1), along with the representation (2) and potential type (3). Next, more simulation parameters (4), such as allowed bonds and the number of heavy atoms, are defined, along with the number of MC steps, moves, and the set of $\beta_{i}$. The $\beta_i$ values were taken to be the same as the ones used in numerical experiments of Ref.~\citenum{mosaics}, \emph{i.e.} we used 4 greedy replicas and 32 exploration replicas, with the latters' $\beta_{i}$ forming a geometric progression between 1.0 and 8.0, which is a rule often used in parallel tempering.\cite{Rathore_Pablo:2005,Kone_Kofke:2005} The curvature value $V_{0}=0.05$ of the potential was selected such that the replica with the highest $\beta_i$ enabled easy global exploration. Finally, MC sampling is performed as outlined in the previous section.

\section{Results}
\label{sec:results}

First, we first visualize and discuss the local clustered chemical space around a few selected molecules and investigate differences in the neighborhood between representations. Then, by generating molecules in spherical shells around central molecules we relate the average slope in chemical space (= distances in properties divided by distances in representation) to molecular property values. In the following text, we use atomic units unless specified.

\subsection{Chemical environment of molecules}

 \begin{figure*}[htb]
          \centering           
          \includegraphics[width=1.0\linewidth]{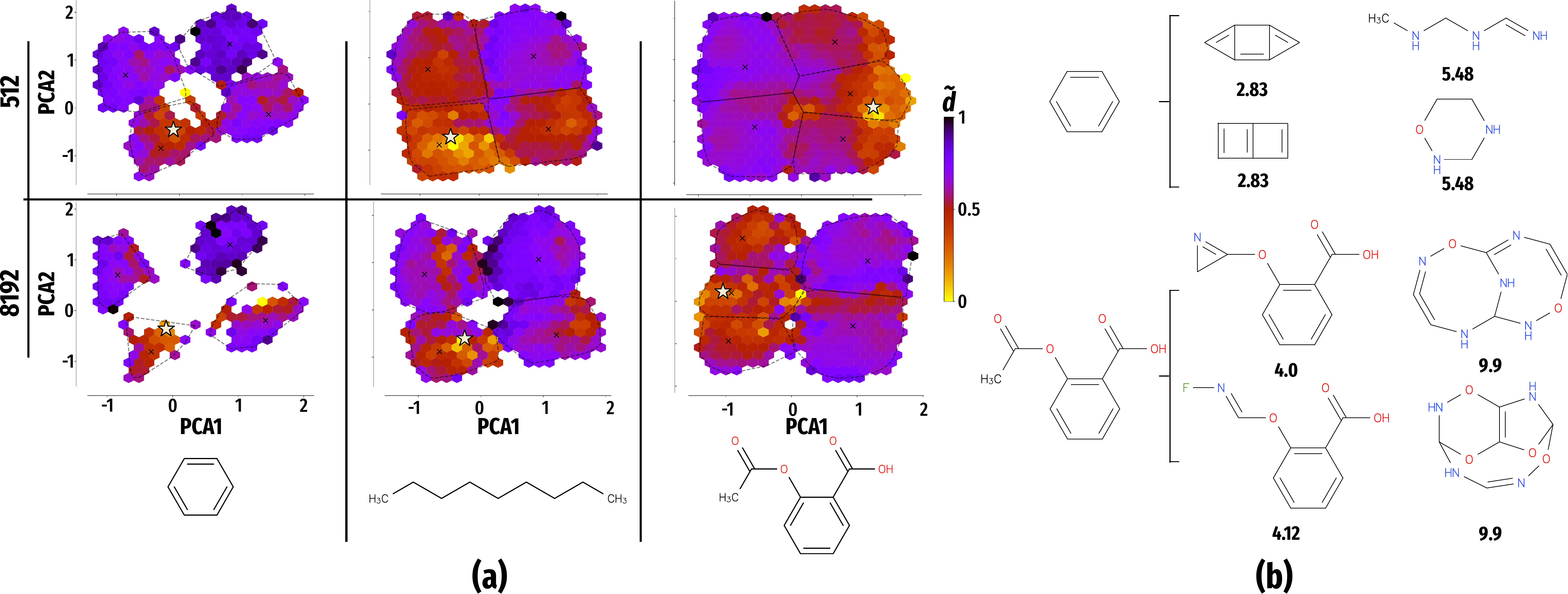}
          \caption{
            (a): Chemical spaces with a fixed number of heavy atoms: benzene six, nonane nine, and aspirin thirteen. We show density plots of the first two principal components (PCA1 and PCA2) obtained from the sampled chemical space. The color gradient denotes the distance $\tilde{d}$ (normalized for maximum distance for each $\nbits$) to the initial molecule in each fingerprint (FP) dimension. Central molecules shown in the bottom panes are represented as white stars in the PCA plots. The upper panel for bit-length $\nbits=512$, lower panel for $\nbits=8192$. Crosses within the visualization indicate the positions of molecules closest to the respective cluster centers, all listed in the SI. (b): closest and farthest molecules encountered for benzene and aspirin respectively using ECFP4 at $\nbits=8192$.}
\label{fig:fp_length_illustration} 
 \end{figure*}
We investigate chemical spaces derived from the QM9 dataset\cite{qm9}, comprising 134k molecules with up to 9 heavy atoms (C, O, N, and F). Primarily, the ``QM9sub'' chemical space is considered, which mirrors the QM9's elemental composition but excludes covalent bonds O-F, O-O, F-F, and N-N due to their high reactivity.

Three central molecules are selected: two from QM9sub and one (aspirin) from QM13sub.
We analyze the surrounding chemical space and via ECFP4 fingerprints and principal component analysis (PCA) for $32 \leq \nbits \leq 8192$ using ECFP4 at $\nbits=2048$ for $d_1=0$, $d_2 = 6$ (see Eq.~(\ref{eq:flat_parabola})). The number of heavy atoms during the simulations is fixed to the number of heavy atoms of the initial molecules \textit{i.e.} exactly six for benzene and thirteen for aspirin. Histograms for the density of the first two principal components of representation vectors for benzene, aspirin, and nonane at $\nbits=512$ and $8192$ are shown in Fig.~\ref{fig:fp_length_illustration}a. Using $k$-nearest neighbor, the optimal cluster count is determined via the elbow method, analyzing explained variation against cluster number. The most representative molecule in each sub-cluster, detailed in SI, is the one nearest to the $k$-means centers (cf. Fig.~\ref{fig:fp_length_illustration}a).

Next, we examine ECFP4 across various $\nbits$ values. For $\nbits=8192$, distinct clusters are observed for benzene and nonane. Benzene's clusters are entirely disconnected, indicating regions within the convex hull of all clusters with varying densities of valid molecular graphs. Reducing $\nbits$ from 8192 to 512 merges distinct clusters and shortens distances between their centers, showing the impact of $\nbits$ on separating chemical environments.

Fig.~\ref{fig:fp_length_illustration}b displays the molecules whose ECFP4 representation vectors are nearest and furthest from those of benzene and aspirin. For aspirin, the closest molecules deviate in the ester group but keep the ring structure. For benzene, the closest points are also hexagonal hydrocarbons with added bonds, while distant molecules include linear structures and varying compositions. Notably, even for $\nbits=8192$ ECFP4 can yield similar or identical distances for substantially different molecular graphs.

\subsection{Differences between representations}

Next, we generate molecules with similar vectors for QM9sub molecules across smaller shell intervals. We select 52 QM9sub molecules as central points and perform $N=5000$ MC steps for ECFP4 and $N=2000$ for SOAP (limited by the higher cost of conformer generation), allowing up to nine heavy atoms. All properties were computed as outlined in Ref.~\citenum{mosaics}. The solvation free energy $G$ and band gap $\varepsilon$ of these molecules are calculated with GFN2-xTB\cite{Bannwarth_Grimme:2019} and a linearized Poisson-Boltzmann model\cite{Ehlert_Grimme:2021}, simulating water presence.

\begin{figure}[htb]
          \centering           
    \includegraphics[width=0.8\columnwidth]{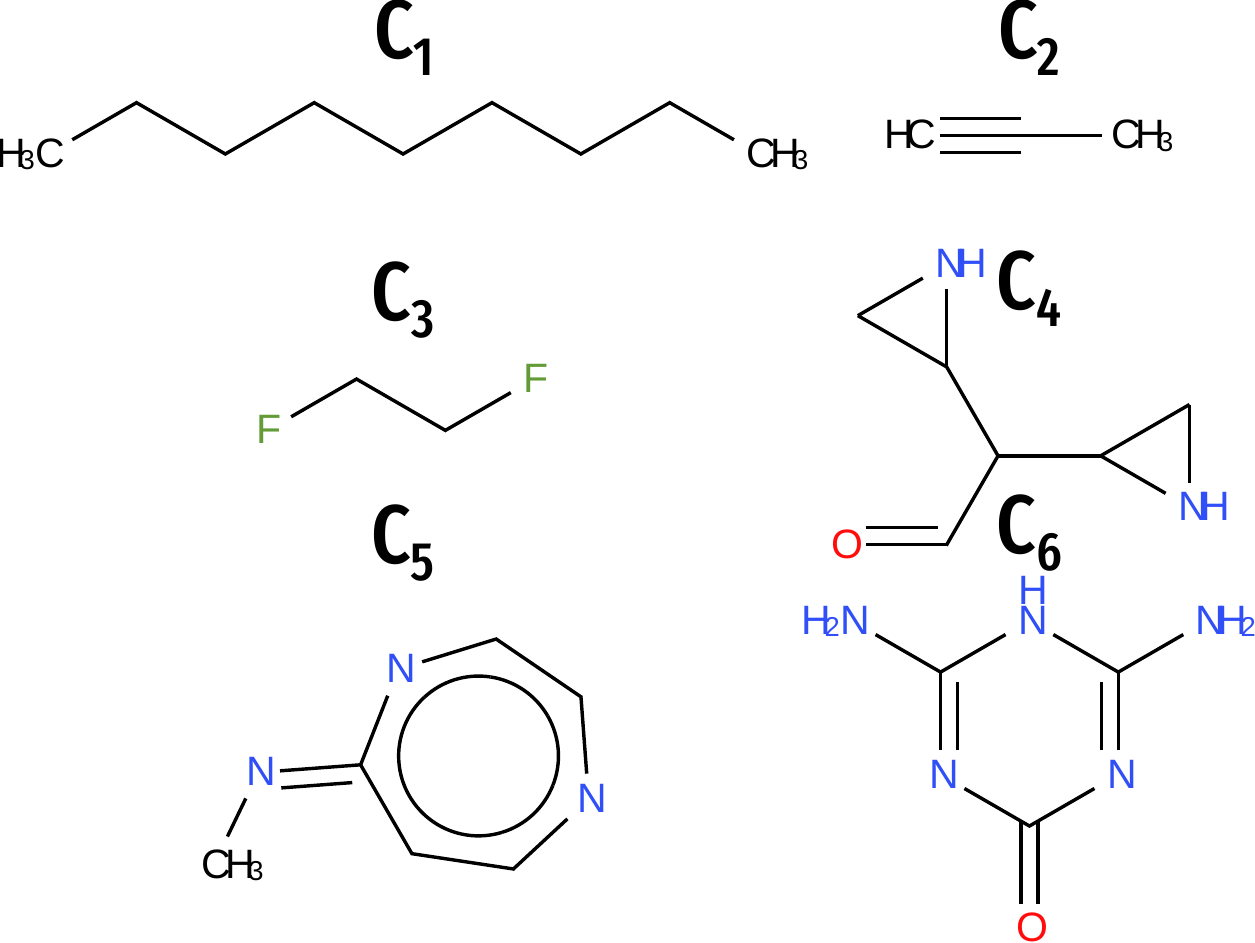}
          \caption{Identified central molecules $\mathbf{C}_l$ with extremal radial average property slope values cf. Tab.~\ref{tab:slopes_and_spearman}.}
     \label{fig:compounds} 
 \end{figure}

SOAP showed better clustering around initial molecules compared to ECFP4, which does not need geometrically constrained graphs (see Fig~1 in SI). The number of molecules near a central molecule varies with its chemical graph, indicating diverse molecular densities and topologies in chemical space. For example, ECFP4 identified 574 unique molecules near ethanol within $2.75\leq d_{\text{ECFP4}} \leq 3.25$, but none for \texttt{NC1=NC(=O)N=C(N)N1} in that range (see $\textbf{C}_{6}$ in Fig.~\ref{fig:compounds}). By contrast, for SOAP and the interval $50\leq d_{\text{SOAP}} \leq 60$, no other molecule is found for ethanol, but for \texttt{NC1=NC(=O)N=C(N)N1}, we find 21 molecules suggesting that the density of molecules around central molecules is highly dependent on the representation type.

\subsection{Property derivatives in chemical space}
\label{sec:relating_d_and_p}
\begin{figure*}[htb]
          \centering           
        \includegraphics[width=\linewidth]{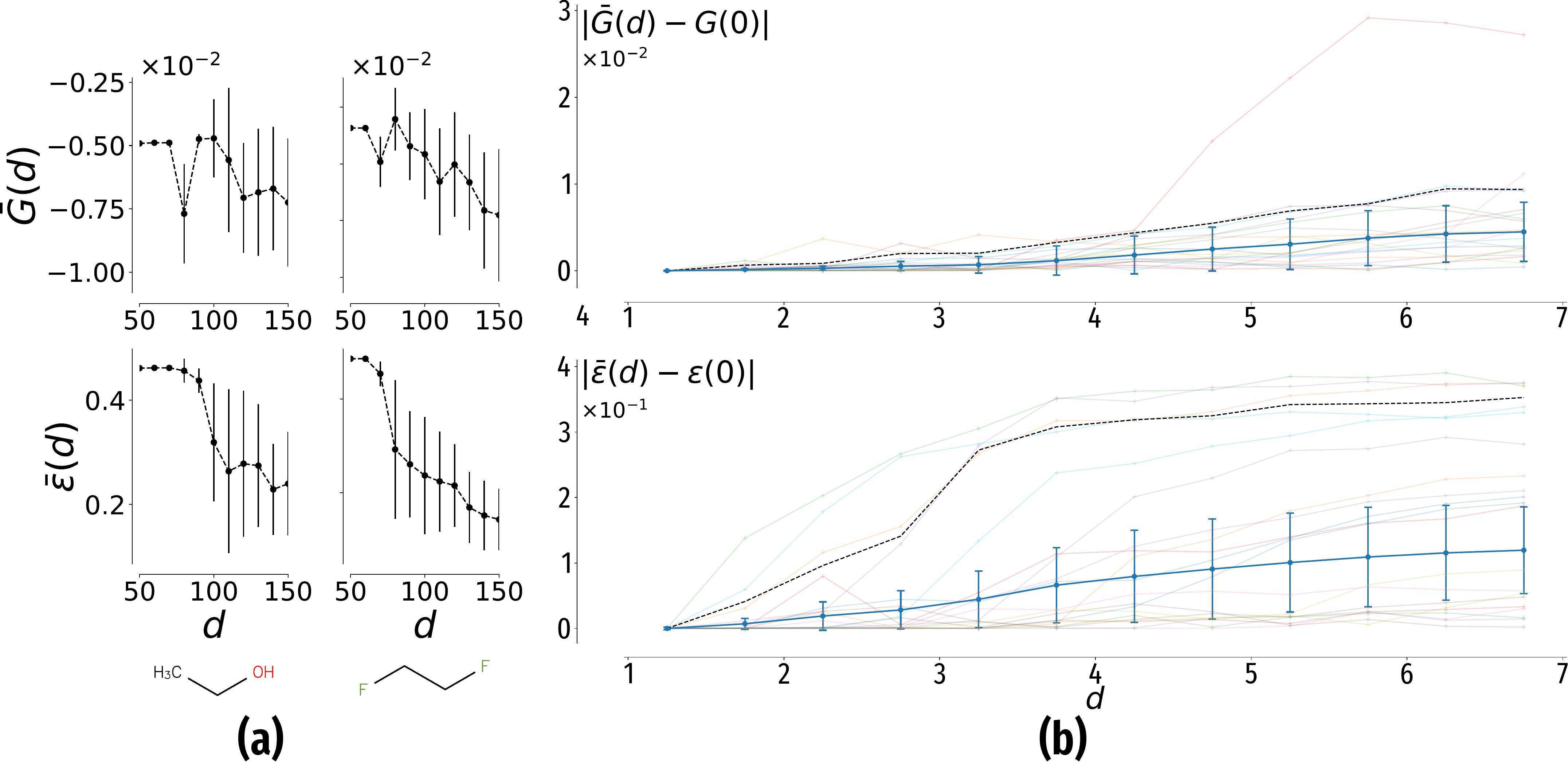}
          \caption{
          (a): radially averaged solvation energy $\bar{G}(d)$ and HOMO-LUMO gap $
          \bar{\varepsilon}(d)$ as a function of SOAP distance $d$ for molecules shown in the bottom panel. (b): Absolute values of radially averaged property shifts $|\overline{P}(d) - P(0) |$ compared to the values $P(0)$ of several initial molecules (opaque) for increasing ECFP4 distances $d$. Black dashed lines show respective $95^\text{th}$ percentiles. The blue line with error bars shows the average shift overall considered molecules at each $d$. top: $P=G$, solvation energy, bottom $P=\varepsilon$, HOMO-LUMO gap.}
     \label{fig:properties_distance} 
 \end{figure*}
By employing exact MC sampling of the local chemical space
-- the space of molecules whose distance to a given molecule is within a threshold -- using distance measures, as elaborated in section \ref{sec:details}, we can now efficiently compute property averages in CCS, a task that was previously unattainable. This method involves calculating averages over specified distance intervals, forming shells as shown in Fig.\ref{fig:radar}, to assess these averages in CCS. Shells are defined by introducing equivalent bins with respect to the distance $d$ to the central molecule where the bin $b$ positions are given by
\begin{align}
     d_{\text{ECFP4}}^{b} = \{1.25+b\cdot 0.5 \}~,
\end{align} 
where $b \in 0...,10$. This results in discrete \emph{shell} intervals of
\begin{align}
  I_{\text{ECFP4}} = \{ [0.75,1.25 ], \underbrace{(1.25,1.75]}_{b=1}, ...,(5.75,6.25 ]  \}
\end{align}
for ECFP4 and for SOAP
\begin{align}
    d_{\text{SOAP}}^{b} = \{50+b\cdot 10 \},
\end{align}
where $b \in 0,...,10$. The shell intervals $I_{\text{SOAP}}$ are defined analogously. The distances do not start at zero as we could not find any molecules for smaller distances than those used here. The bin centers are chosen according to meaningful distances for each representation respectively based on preliminary simulations.
For instance, at a SOAP distance of $d=50$, only a single bond may differ from the central molecule, whereas at $d=150$ encountered molecules will not at all resemble the central molecule and similar for ECFP4. However, a more systematic approach to defining these intervals, such as using average molecular distances in the relevant chemical space as a reference, is worth exploring. For all 52 central molecules, with index $l$, the average of both properties within each finite shell interval is computed and denoted as $\bar{P}_{l}(d)$ (the property $P$ can be $G$ or $\varepsilon$), \emph{e.g.} $\bar{G}(d=1.25)$ is the average value of free energy of solvation over all molecules in the interval $[0.75, 1.25]$ for ECFP4. More generally, we define shell averages as
\begin{align}
    \bar{P}_{l}(d^{b}) = \frac{1}{M_b} \sum_{j=1}^{M_{b}} P_{k},
    \label{eq:def_shell_average}
\end{align}
where $M_{b}$ is the number of molecules in shell $b$. These are molecules whose distance to the central molecule $l$ is between $d^{b-1} $ and $ d^{b}$. The indices $j\in 1, ..., M_{b}$ enumerate these molecules. In this notation the property value of the central molecule $l$ is $\bar{P}_{l}(0)$. For simplicity the index $b$ representing the shell index is removed.

Considering the number of encountered molecules grows quickly with distance we introduce a cutoff, evaluating only up to 300 randomly selected molecules per shell to limit computational costs. As shown in Fig.~\ref{fig:properties_distance}a for small distances, $d_{\text{SOAP}}\leq 60$, property values of new molecules are similar to those of the two central molecules. Increasing the distance further leads to a larger drift in the property values -- consistent with the assumption of regression models that small distances $d$ between training and test molecules correlate to similar $P$.
Both $\bar{G}(d)$ and $\bar{\varepsilon}(d)$ vary substantially from molecule to molecule, as shown in Fig.~\ref{fig:properties_distance}b making it challenging to extract general trends for $\bar{P}(d)$. Still, we fit linear functions to $\bar{P}_{l}(d)$ to approximate average trends of properties through chemical space for each molecule $l$ as follows,
\begin{equation}
\bar{P}_{l}(d) \approx m^{P}_{l} \cdot d + c^{P}_{l},
\label{eq:property_linear}
\end{equation}
where $m^{P}_{l}$ and $c^{P}_{l}$ are the fitted slope and intercept.
 \begin{figure*}[htb]
          \centering           
        \includegraphics[width=0.8\linewidth]{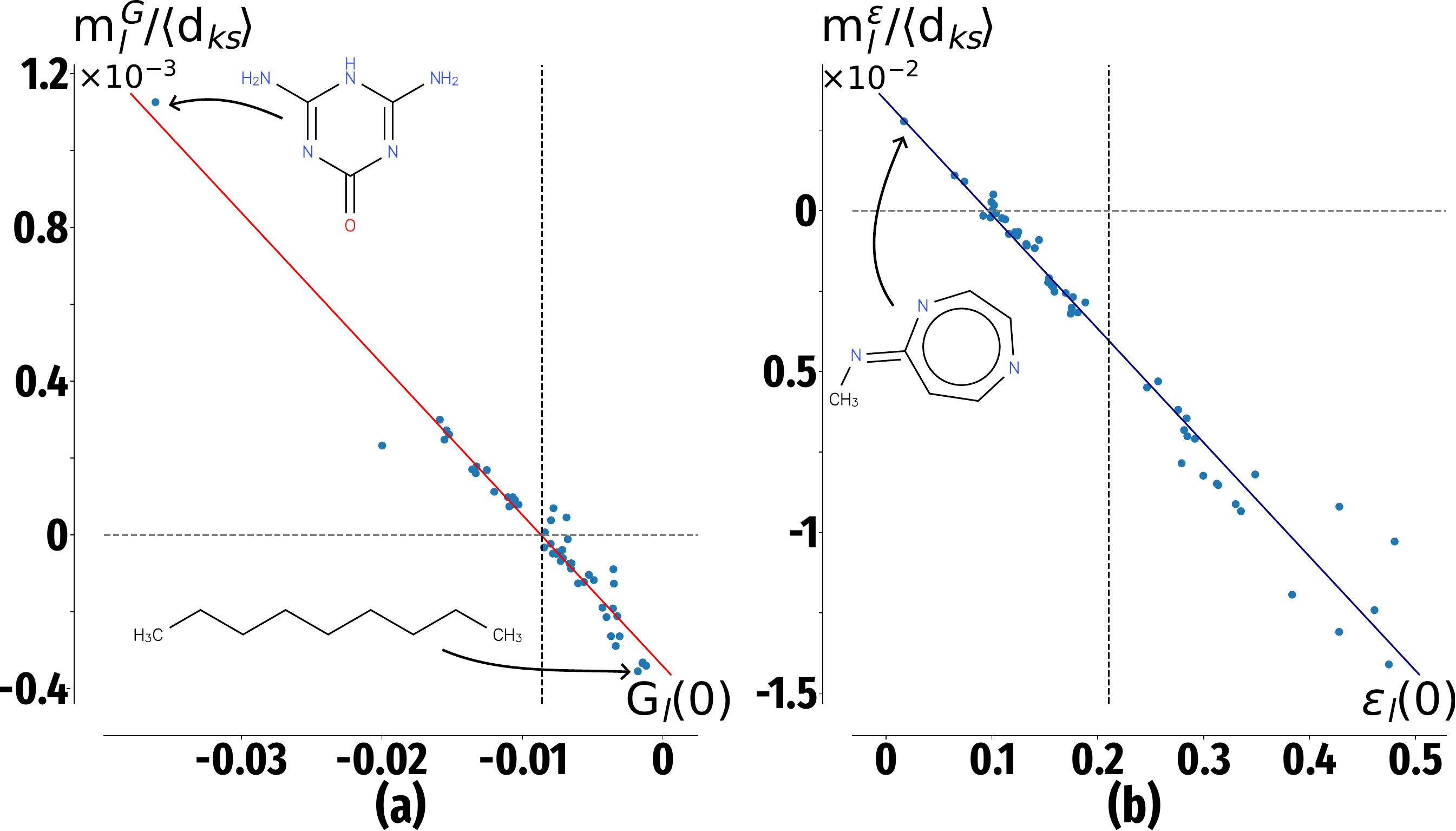}
          \caption{
          Slopes $m^{P}_{l}/\langle d_{ks} \rangle$ of average changes of a property [(a) free energy of solvation $G$ or (b) band gap $\epsilon$] with respect to $d$ for molecules surrounding a central molecule using ECFP4 (SOAP shown in the SI Fig.~1, and values given in Tab.~\ref{tab:slopes_and_spearman}). 
          The slope values are scaled by the average pairwise distance $\langle d_{ks} \rangle$ of all 52 molecules.
          Each dot corresponds to a molecule $l$. y-axis displays the slope of a linear function fitted to Eq.~(\ref{eq:property_linear}) for each property. x-axis represents the central molecule's property. Molecules with the largest and smallest slopes are shown in the inset. Top: solvation energy $G$, bottom HOMO-LUMO gap. The vertical black dashed line indicates the property average over all 52 central molecules and the horizontal line indicates zero slope.}
     \label{fig:slope_properties} 
 \end{figure*}
After extracting $\bar{\varepsilon}_{l}(d)$ and $\bar{G}_{l}(d)$ for each molecule and each $d$, we compute the slope values $m^{G}_{l}$ and $m^{\varepsilon}_{l}$ by regressing Eq.~(\ref{eq:property_linear}) to the outer shell values as $d$ and the 10 shell averages $\bar{P}_{l}(d)$. Note that $c^{P}_{l}$ is determined by the value of the central molecule. Scatter plots displaying the slope values $m^{P}_{l}$, normalized by the average pairwise distance of all central molecules $\langle d_{ks} \rangle$, against the property values of the central molecules ($d=0$) are presented in Fig.~\ref{fig:slope_properties}a,b. 

A near linear correlation between the radially averaged property $m_{l}^{P}$ derivative and the central property value is found, $m_{l}^{P} \propto \bar{P}_{l}(0)$. For instance, the extracted slope values for the solvation free energies $m_{l}^{G}$ correlate linearly with the initial value of the respective central molecule $G_{l}(0)$. Average property values increase (decrease) for molecules from the tails of the property distribution corresponding to smaller (larger) values for increasing shell distances. Radial average property values of sampled molecules from initial compounds with more negative solvation energies than most other molecules drift towards less soluble compounds and \textit{vice versa}. The same trend applies to the HOMO-LUMO gap $\varepsilon$. We also find that central molecules with average property values have a near zero slope, indicated by vertical and horizontal lines in Figs.~\ref{fig:slope_properties}a and~\ref{fig:slope_properties}b respectively.

The linear trend holds for the band gap (intensive) and solvation energy (extensive) quantity independently as we cannot find a correlation between both quantities (SI Fig.~4b). It should be noted though that we have used the the average SOAP representation (intensive) instead of the sum of SOAP terms (extensive). It would therefore be interesting to investigate further how the observed trends behave with explicitly size-extensive global representations\cite{https://doi.org/10.1002/syst.201900052}.

Given $m^{P}_{l} \propto \bar{P}_{l}(0)$, an upper bound of the average property slope $\bar{P}^{\prime} = m^{P}$ in a chemical space can be estimated by the compound with the largest property value,
\begin{align}
 \max\bar{P}^{\prime}(d)=\max  m^P  =\max\limits_{l \in \text{CCS}}\left[ \slopeofslopes \bar{P}_{l}(0) + \interceptofslopes\right],
 \label{eq:max_slope}
\end{align}
where $\slopeofslopes$ and $\interceptofslopes$ are linear slope and intercept extracted from the correlation plots in Figs.~\ref{fig:slope_properties}a and~\ref{fig:slope_properties}b. The most negative slope on the other hand is defined by the molecule with the minimal property value. The regression constants $\slopeofslopes$, $\interceptofslopes$ as well as the average, minimal, and maximal slope values and Spearman's rank correlation coefficient $r_s$, measuring monotonic correlation between $\bar{P}^{\prime}$ and $\Delta_{l}^{P} $, for both properties and representations are listed in Tab.~\ref{tab:slopes_and_spearman}. The identified extremal molecules are shown in Fig.~\ref{fig:compounds}. As expected $\slopeofslopes$ and $\interceptofslopes$ depend on the property, representation, and permitted CCS. For both representations, the magnitude of the maximal slope for $\bar{G}$ corresponds to the same compound $\mathbf{C}_{6}$ (s. Fig.~\ref{fig:compounds}) which is also the most soluble in QM9. The correlation coefficient $r_s$ for the SOAP representation is smaller than that of the ECFP4 fingerprint. This is evident from the values recorded for $G$, which are $-0.788$ and $-0.976$ respectively, as shown in Tab.~\ref{tab:slopes_and_spearman}. We propose two potential explanations for the higher correlation coefficient of ECFP4.
First, we managed to perform more MC steps for ECFP4 simulations, thus potentially making them more converged. Second, the conformer sampling as required for the SOAP workflow introduces statistical noise due to a degree of randomness in generating conformers used in SOAP representation.

\begin{table*}[htb]
    \centering
    \caption{Minimal and maximal slope $m^P$ of the average radial property derivative in chemical space and corresponding molecules 
    $\mathbf{C}_i$ identified among 52 QM9sub molecules as well as the Spearman correlation coefficient, shown in Fig.~\ref{fig:compounds}. Results for both molecular representations, ECFP4\cite{ecfp} as well as SOAP.\cite{soap}}
    \label{tab:slopes_and_spearman}
    \begin{tabular}{lcccccccc}
                            & $P$                               & Min mol & Max mol & $\text{min}~  m^P$      & $\text{max}~m^P$       & $\slopeofslopes$ & $\interceptofslopes$ & $r_{s}$ \\
        \hline
        \multirow{2}{*}{ECFP4} & $G$                               & $\textbf{C}_{1}$ & $\textbf{C}_6$ & $-2.02 \times 10^{-3} $ & $6.42 \times 10^{-3} $ & $-2.24 \times 10^{-1}$ & $ 1.94 \times 10^{-3}$ & -0.976 \\
                               & $\varepsilon$                     & $\textbf{C}_1$ & $\textbf{C}_5$ & $-8.04 \times 10^{-2} $ & $1.58 \times 10^{-2} $ & $2.02 \times 10^{-1}$ & $ 1.94\times 10^{-2}$ & -0.988 \\
        \hline
        \multirow{2}{*}{SOAP}  & $G$                               & $\textbf{C}_2$ & $\textbf{C}_6$ & $-7.32 \times 10^{-5} $ & $8.74 \times 10^{-5} $ & $-5.24 \times 10^{-3}$ & $4.66 \times 10^{-5}$ & -0.788 \\
                               & \multicolumn{1}{l}{$\varepsilon$} & $\textbf{C}_3$& $\textbf{C}_4$ & $-3.41 \times 10^{-3} $ & $4.99 \times 10^{-4} $ & $ -5.46\times 10^{-3}$ & $-7.47 \times 10^{-4}$ & -0.874 \\
    \end{tabular}
\end{table*}

Since Eq.~(\ref{eq:max_slope}) describes a relation over all of the permitted chemical space, even approximate values for the maximal $m^{P}$ may help design molecular representations or tuning model flexibility by taking into account distances between data points and the expected variation in the target property. As an illustration, we show the absolute property deviation from the initial values in Figs.~\ref{fig:properties_distance}a and ~\ref{fig:properties_distance}b: A too small slope of the curve under the $95^\text{th}$ percentile means weak correlation with the target property. Larger slopes on the other hand do not necessarily lead to smaller errors as they can correspond to discontinuities of the representation with respect to the property, which is associated with larger errors.\cite{D2MA00742H}

\section{Conclusion}
\label{sec:conclusion}

An MC-based method was introduced to enable exact sampling of distributions of molecules defined in terms of distances within the original feature space. 
Essentially, the sampling probabilities are determined by a comparison of molecular features. This is particularly useful for interpretable descriptors commonly used in cheminformatics and may help in analyzing feature-based representation vectors. Using the introduced sampling method a nearly linear trend between the initial value of the central molecule and the radial slope, $m^{P}_{l} \propto \bar{P}_{l}(0)$, was found (s.~Fig.~\ref{fig:slope_properties}). 
Note that the necessary simulations (s.~\ref{sec:relating_d_and_p}) were only possible because of exact sampling with respect to the distances. Using the linear relationship a lower bound for the maximal slope in chemical space can be estimated given by the molecule with the largest property value [s. Eq.~(\ref{eq:max_slope})]. We hypothesize this trend is due to the difference between the initial molecule and the average property value, but this observation warrants further investigation. Several aspects could be investigated building on this work: (i) incorporating control for synthetic accessibility\cite{doi:10.1021/acs.jcim.2c00246} into the generation of molecules will further enhance its utility. 
(ii) Identifying pairs of molecules that are close in representation but have significantly different properties, sometimes referred to as activity cliffs,\cite{doi:10.1021/acs.jcim.2c01073} presents a challenge for existing predictive models -- in particular for ligand-based drug-design applications of ML. (iii) Investigating the effect of different choices of metrics on the structure of CCS, for instance, Tanimoto\cite{Tanimoto1958} or Wasserstein.\cite{wasserstein,10.1063/1.4964627} (iv) Systematic generation of molecules far from an initial molecule may find application for farthest point sampling often used for active learning strategies.\cite{DRAL2020291,doi:10.1021/acs.jpcc.2c03854}
Finally, the method presented here may also serve as a tool to invert molecular representations with applications to molecular property optimization.

\section{Supplementary Information}

The following Supplementary Information contains the SOAP representation figure corresponding to Fig.~\ref{fig:slope_properties} and the PCA corresponding to the solvation-free energies.

\section{Code Availability}

The Python code and graphical user interface for the \texttt{ChemSpaceSampler} is available at 
\url{https://github.com/chemspacelab/mosaics}. The notebooks with examples can be found here
\url{https://github.com/chemspacelab/mosaics/tree/main/examples/05_chemspacesampler}.

\section{Implementation}

As illustrated in Fig.~\ref{fig:components_code} usage steps for the \texttt{ChemSpaceSampler} code are as follows: (1) define the central molecule, (2) representation, (3) potential type, and (4) constraints for the chemical space, such as permitted bonds, number of heavy atoms, along with simulations details such as number of MC steps, moves, and set of $\beta_{i}$. In practice, using the code is as simple as defining a set of parameters and calling the main function as follows:
\begin{lstlisting}[language=Python]
params = {
    'min_d': 2.0,
    'max_d': 5.0,
    'Nsteps': 500,
    'possible_elements': ["C", "O", "N", "F"],
    'forbidden_bonds': [(8, 9), (8, 8), (9, 9), (7, 7)],
    'nhatoms_range': [1,9],
    'betas': [8, 4, 2],
    "nBits": 2048,
    'rep_name': 'ECFP'
          }

MOLS, D = chemspace_potentials.chemspacesampler_ECFP(smiles="CC(=O)OC1=CC=CC=C1C(=O)O", params=params)    
\end{lstlisting}
The parameter dictionary comprises setting the desired minimal and maximal distance of the constant potential, the number of MC steps, and possible elements. It includes the option of excluding certain bond combinations. In addition, the number of bits $\nbits=$ \texttt{nBits}, inverse temperature, and representation type are defined. Finally, the simulation is initialized with a SMILES string returning the encountered molecules and their distances from the central molecule.

\section{Data Availibility}

The data that support the findings of this study are available from the corresponding author upon request.

\section{Acknowledgments}

We acknowledge discussions, proofreading, and valuable scientific input which have been essential to this project from Prof. Dr. Guido Falk von Rudorff. This project has received funding from the European Union’s Horizon 2020 research and innovation programme under grant agreement No 957189. Obtaining the presented computational results has been facilitated using the queueing system implemented at \href{https://leruli.com}{http://leruli.com}. J.W. acknowledges support from the Faculty of Physics and supervision by Anatole v. Lilienfeld and C. Dellago at the University of Vienna. The authors thank Michael Sahre, Alexandre Schoepfer and Matthieu Haeberle for their helpful discussions and feedback.

\bibliography{literature}

\end{document}


\title[]{Exact sampling of molecules in chemical space} 
\newcommand{\nbits}{n_{\mathrm{bits}}}
\author{Jan Weinreich}

\affiliation{University of Vienna, Faculty of Physics, Kolingasse 14-16, AT-1090 Vienna, Austria}
\affiliation{University of Vienna, Vienna Doctoral School in Physics, Boltzmanngasse 5, 1090 Vienna, Austria}
\author{Konstantin Karandashev}
\affiliation{University of Vienna, Faculty of Physics, Kolingasse 14-16, AT-1090 Wien, Austria}
\author{Guido Falk von Rudorff}
\affiliation{University Kassel, Department of Chemistry, Heinrich-Plett-Str.40, 34132 Kassel, Germany}
\alsoaffiliation{Center for Interdisciplinary Nanostructure Science and Technology (CINSaT), Heinrich-Plett-Straße 40, 34132 Kassel}
\email{vonrudorff@uni-kassel.de}

\date{\today}

\maketitle 

\bigskip

\subsection{Diversity of Chemical Space}

\begin{figure*}[htb]
          \centering           
          \includegraphics[width=0.8\linewidth]{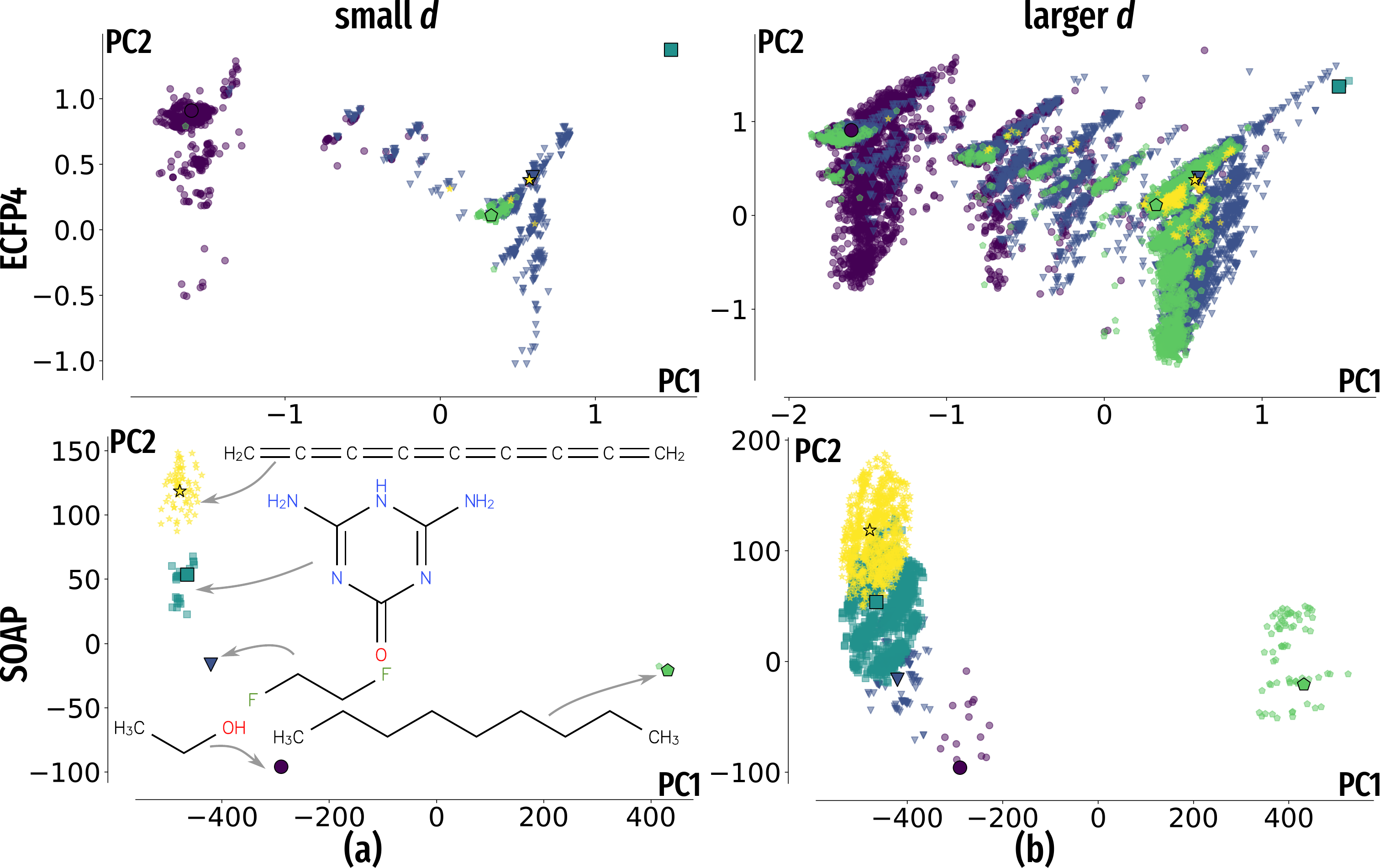}
          \caption{
          \texttt{ChemSpaceSampler} with ECFP4 (top) and SOAP\cite{soap} (bottom) representations: comparing sampled chemical space projected principal components (PC) into two dimensions along axis PC1 and PC2 with maximal variance. Large marker symbols represent central molecules used in the potential function, Eq.~(2) in main text. Small opaque markers with matching colors and styles denote molecules obtained after sampling. (a) distance intervals at small representation distances: $2.75\leq d_{\text{ECFP4}} \leq 3.25$ for ECFP4 and $50\leq d_{\text{SOAP}} \leq 60$ for SOAP; (b) same at larger distances: $4.00 \leq d_{\text{ECFP4}} \leq 4.25$ and $100\leq d_{\text{SOAP}} \leq 110$.}
     \label{fig:compare_local_sampling} 
 \end{figure*}

\begin{figure*}[htb]
\centering
\includegraphics[width=0.8\linewidth]{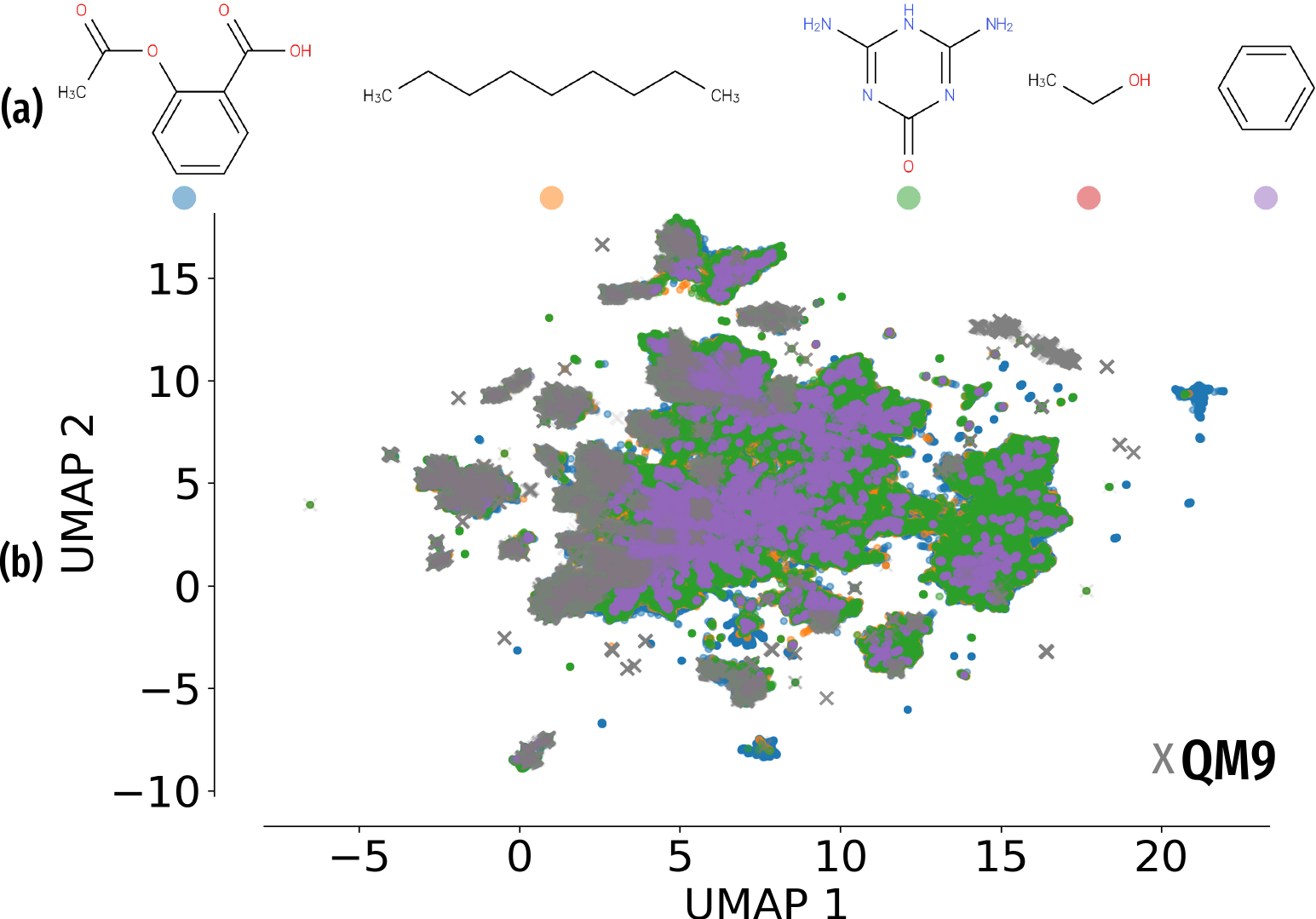}
\caption{UMAP projection of the QM9sub chemical space explored by the \texttt{ChemSpaceSampler}. Sampling the local chemical space of five central molecules results in a diverse range of molecular structures, as illustrated by the dispersion of points throughout the plot: For instance, purple dots correspond to molecules found by \texttt{ChemSpaceSampler} using benzene as a central molecule. Grey crosses represent molecules from the QM9 dataset.}
\label{fig:umap_chemical_universe}
\end{figure*}

We sample CCS of five randomly selected central molecules (shown in s. Fig.~\ref{fig:umap_chemical_universe}a bottom) of which four are from the QM9sub space with ECFP4 at $\nbits=2048$ for $\gamma=0$, $\sigma = 6$, see Eq.~(2) in main text. In addition, we study aspirin, from QM13sub with 13 heavy atoms. A comparison between the QM9\cite{qm9} dataset (grey) and the \texttt{ChemSpaceSampler} QM9sub results is shown as a two-dimensional UMAP\cite{McInnes2018} visualization in Fig.~\ref{fig:umap_chemical_universe}b. A diverse chemical space is explored spanning regions not covered by QM9.\cite{qm9} Sampling the chemical neighborhood of only five distinct molecules is sufficient to span a chemical space much larger than the original QM9 dataset -- even though the sampled QM9sub space is more restricted due to additional bond constraints. This demonstrates that even imposing proximity to a given molecule allows generation of chemically diverse sets of molecules.

\subsection{Linear correlation between properties and slope in chemical space}

\begin{figure}[htb]
    \centering           
    \includegraphics[width=0.9\linewidth]{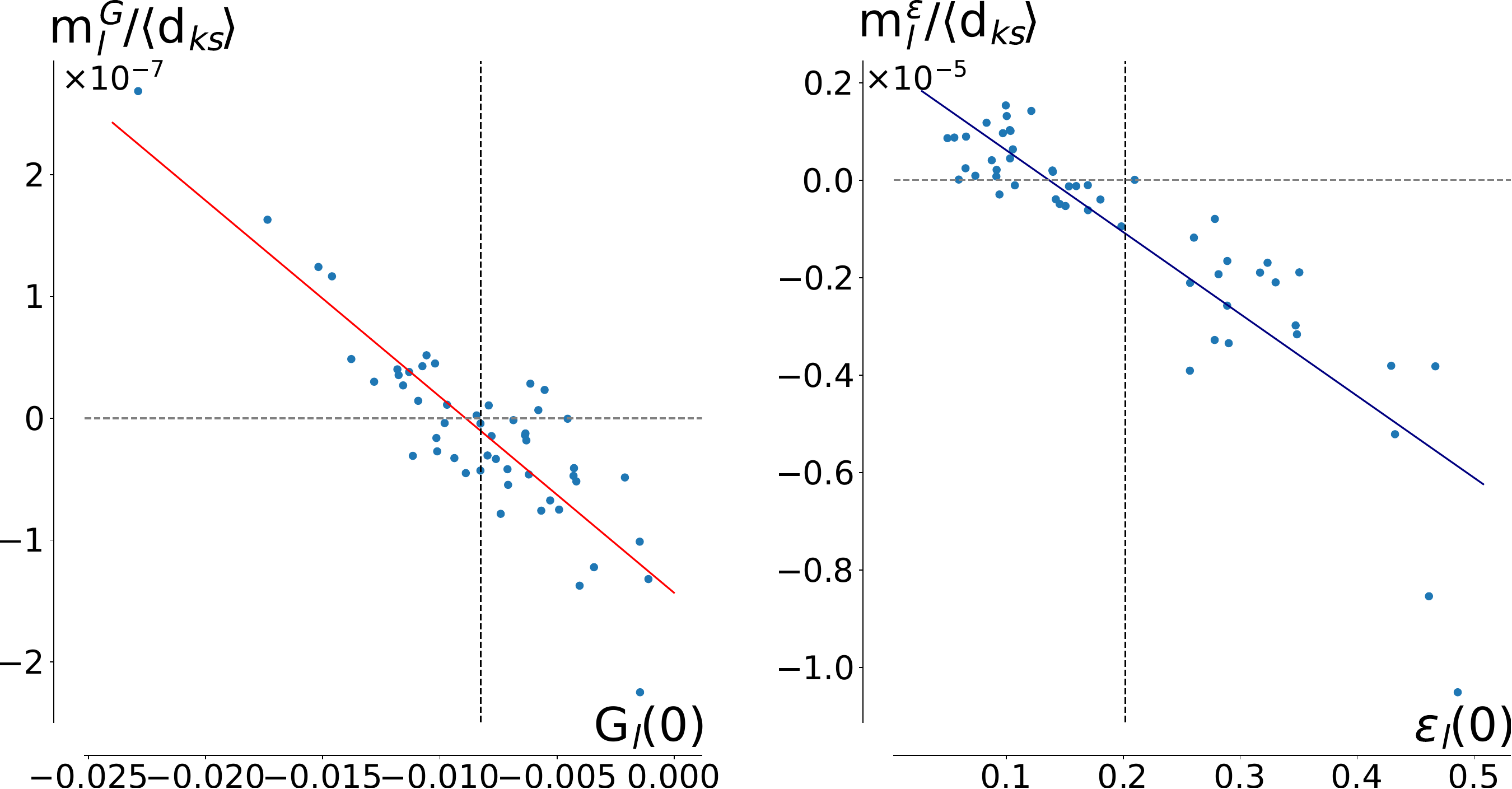}
    \caption{
    For the SOAP representation: Slope $m^{P}_{i}$ of the average property with respect to the representation distance $d$ for molecules surrounding a central molecule, with values given in Table 1 in the main text. Each dot corresponds to a molecule $l$. The y-axis displays the slope of a linear function fitted to each property, and the x-axis represents the central molecule $l$ property value. Left: solvation energy $G$, right: HOMO-LUMO gap (both in atomic units).}
    \label{fig:slope3d}
\end{figure}

Here, we show the corresponding figure to Fig. 6 from the main text but using the SOAP\cite{soap} representation (see Fig.~\ref{fig:slope3d}a,b). As in the corresponding plot in the main text that used the ECFP representation, we find a good linear correlation between the property of the central molecule (solvation energy $G$ and HOMO-LUMO gap $\epsilon$).

\subsection{Lack of correlation between band-gap and solvation energies}
 
\begin{figure*}[htb]
          \centering           
    \includegraphics[width=0.7\linewidth]{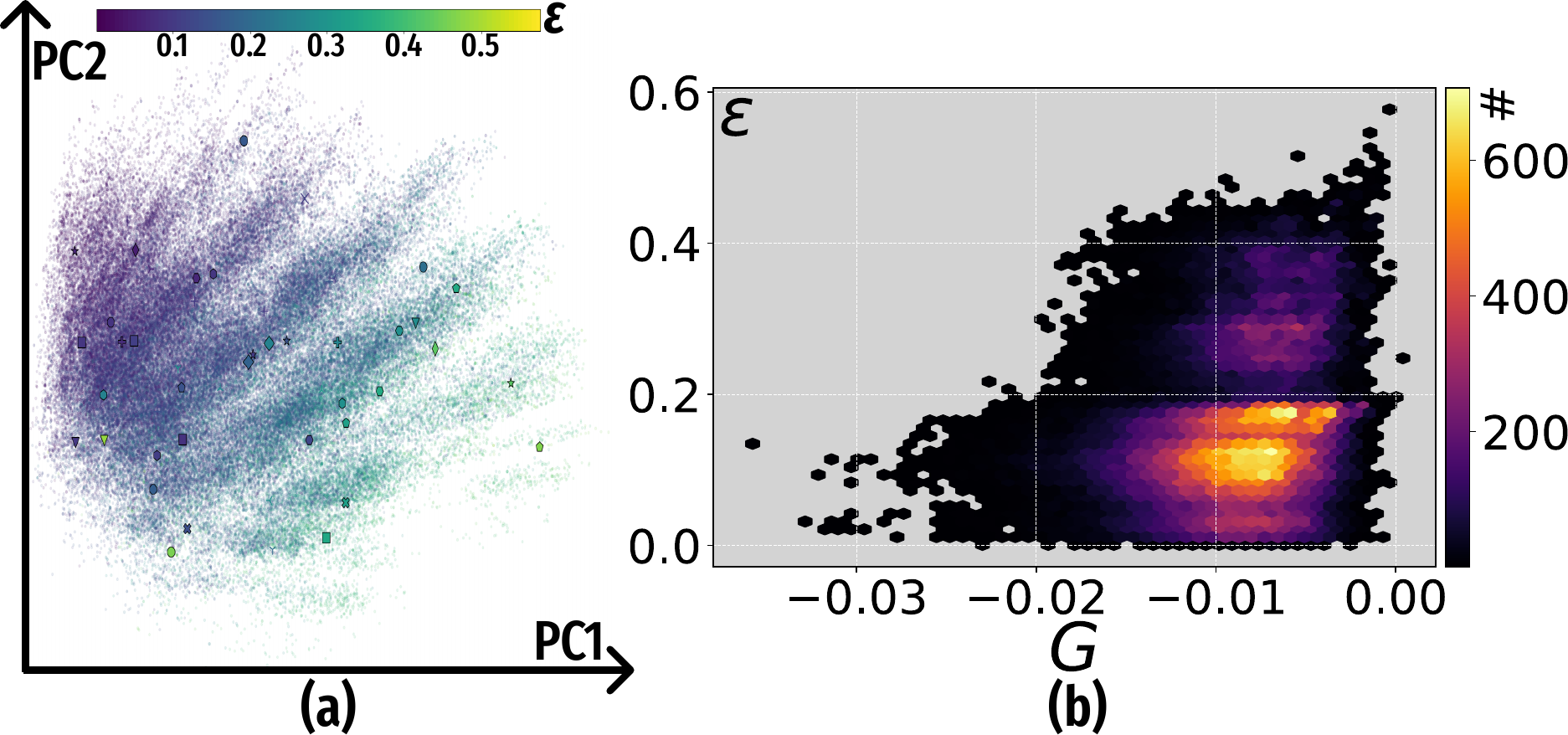}
          \caption{
          (a): first two principal components (PC) of central molecules, shown with markers, as well as sampled molecules, visited in all shells $d \in \bigl[  50, 150 \bigr] $ represented as dots colored by individual HOMO-LUMO gap values $\varepsilon$. (b): scatter plot between solvation energy $G$ and band gap $\varepsilon$ for all molecules obtained by sampling. In both cases SOAP was used.}
     \label{fig:non_correlation} 
 \end{figure*}

To demonstrate that the linear trends demonstrated in Fig.~\ref{fig:slope3d}a,b and Fig.~6 from the main text are observed for two properties that do not correlated (otherwise observation of linear trend for one would automatically lead to corresponding observation for the other), we analyze their interrelationship. A scatter plot of the corresponding $G$ and $\varepsilon$ values for all molecules encountered in the simulation that corresponds to the linear correlation is shown in Fig.~\ref{fig:non_correlation}b. However, there is no visible correlation between the two properties.

\begin{figure}[htb]
          \centering           
          \includegraphics[width=0.5\linewidth]{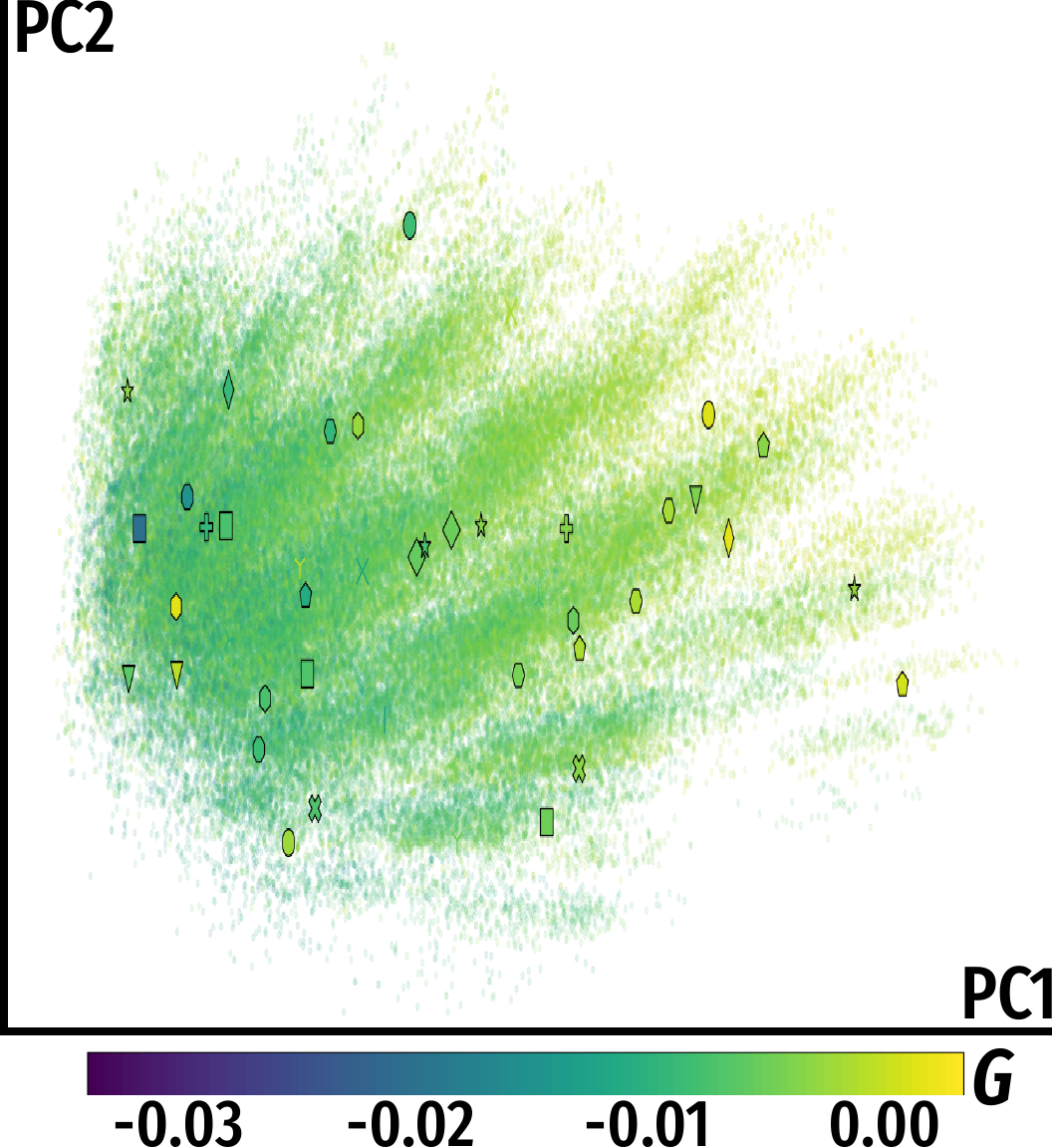}
          \caption{First two principal components of the 50 central molecules, shown with markers, as well as sampled molecules, visited in all shells $d \in \bigl[ 50, 150 \bigr] $ represented as dots colored by individual $G$ values of free energies of solvation in water.}
     \label{fig:pca_soap_G}
\end{figure}

Additionally, the first two principal components of all molecules encountered in the SOAP simulations were collected to illustrate how the HOMO-LUMO gap $\bar{\varepsilon}$ varies across different molecules (see Fig.~\ref{fig:non_correlation}a). The corresponding figure for the solvation energy $G$ is shown in Fig.~\ref{fig:pca_soap_G}. Inspecting the feature space in Fig.~\ref{fig:pca_soap_G}, we can observe a noticeable gradient along the first principal component $\text{PC}1$, where molecules with more negative $G$ tend to have smaller $\text{PC}1$ values, and \textit{vice versa}.

\subsection{Converging distances as a function of resolution in chemical space}
\begin{figure}[htb]
    \centering
\includegraphics[width=0.6\linewidth]{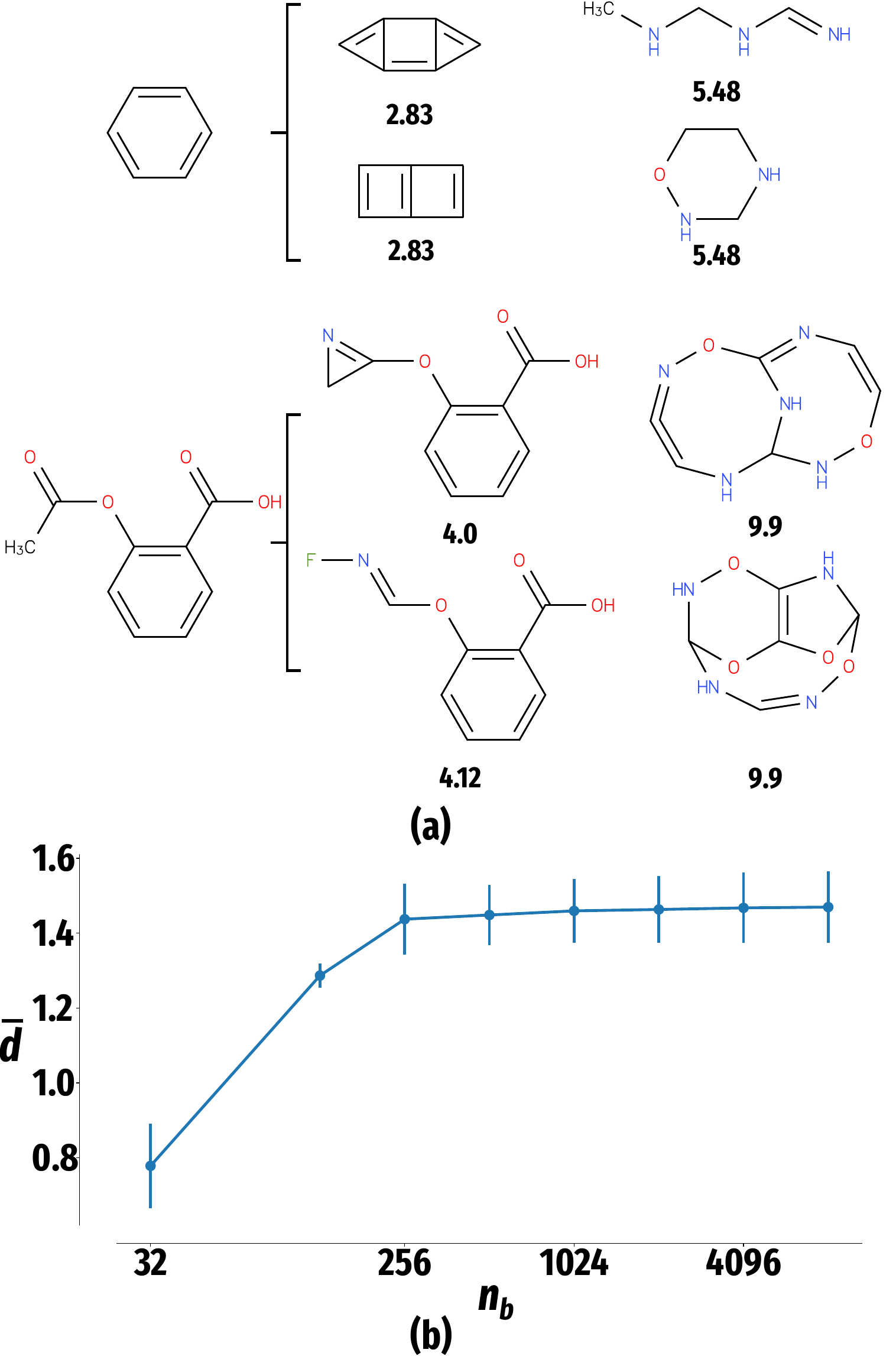}
    \caption{
    (a): closest and farthest molecules encountered for benzene and aspirin respectively using ECFP4 at $\nbits=8192$. (b): mean distance $\bar{d}$ of the clusters averaged over five different molecules, cf. Eq.1.}
    \label{fig:distances_close_far} 
\end{figure}

When clustering molecules in chemical space, the number of clusters will vary depending on the selection threshold that defines a cluster and the specific chemical space, with diverse spaces potentially having more clusters. This suggests the need for a more robust measure to define clustering in CCS. Distinct clustering of parts of chemical space corresponds to the degree of resolution with which different parts of the feature space can be distinguished. To remove the arbitrary dependence on the number of clusters, we evaluate, as a function of fingerprint length, the average distance between molecules closest to the center of each cluster:
\begin{align}
    \bar{d}(\nbits) &=  \frac{1}{M} \sum_{l=1}^{M} \frac{1}{C^{l}} \sum_{{k \neq s}}^{C^{l}} d(A^{l}_{k}, A^{l}_{s})
    \label{eq:dimension}
\end{align}
where $M$ is the number of considered central molecules, $l=1,\dots,M$ is the index of a central molecule, $C^{l}$ is the number of clusters of molecules observed after running an MC simulation, $k$ and $s$ are indices of such clusters, and $A_{k}^{l}$ is the molecule closest to the center of the cluster with index $k$. The distances are averaged over all central molecules to ensure $\bar{d}$ does not depend on the individual chemical diversity around each of them. Intuitively, $\bar{d}(\nbits) \rightarrow 0$ corresponds to the merging of previously separated clusters, whereas a higher resolution leads to larger average distances between clusters. At $\nbits \approx 1024$, the average distance appears to saturate (cf. Fig.~\ref{fig:distances_close_far}b), supporting the choice of the default value $\nbits=1024$ used in RdKit.

\bibliography{literature}